\documentclass[10pt, conference, letterpaper]{IEEEtran}
\IEEEoverridecommandlockouts

\usepackage{cite}
\usepackage{amsmath,amssymb,amsfonts}
\usepackage{algorithmic}
\usepackage{graphicx}
\usepackage{textcomp}
\usepackage{xcolor}
\usepackage{multirow}
\def\BibTeX{{\rm B\kern-.05em{\sc i\kern-.025em b}\kern-.08em
    T\kern-.1667em\lower.7ex\hbox{E}\kern-.125emX}}
\usepackage[linesnumbered,ruled]{algorithm2e}
\SetKwInput{KwInput}{Input}                %
\SetKwInput{KwOutput}{Output}              %
\usepackage{subfig}
\usepackage{cleveref}   

\begin{document}
\bstctlcite{IEEEexample:BSTcontrol}

\title{Janus: Collaborative Vision Transformer Under Dynamic Network Environment\\
}

\author{\IEEEauthorblockN{Linyi Jiang\IEEEauthorrefmark{1},
Silvery D. Fu\IEEEauthorrefmark{2}, Yifei Zhu\IEEEauthorrefmark{1} and
Bo Li\IEEEauthorrefmark{3} \thanks{This work was supported by the National Key R\&D Program of China (Grant No. 2024YFC3017100) and the National Natural Science Foundation of China (Grant No. 62302292). The work of Bo Li was supported in part by a RGC RIF grant under the contract R6021-20, RGC TRS grant under the contract T43-513/23N-2, RGC CRF grants under contracts C7004-22G, C1029-22G and C6015-23G, and RGC GRF grants under the contracts 16200221, 16207922 and 16207423. Corresponding author: Yifei Zhu. }}
\IEEEauthorblockA{\IEEEauthorrefmark{1}UM-SJTU Joint Institute, Shanghai Jiao Tong University\\
\IEEEauthorrefmark{2}Systems Design Studio LLC\\
\IEEEauthorrefmark{3}Department of Computer Science and Engineering, Hong Kong University of Science and Technology\\
Email: jiangly01@sjtu.edu.cn, silvery@sd.studio, yifei.zhu@sjtu.edu.cn, bli@cse.ust.hk}}
\maketitle

\begin{abstract}
Vision Transformers (ViTs) have outperformed traditional Convolutional Neural Network architectures and achieved state-of-the-art results in various computer vision tasks. 
Since ViTs are computationally expensive, the models either have to be pruned to run on resource-limited edge devices only or have to be executed on remote cloud servers after receiving the raw data transmitted over fluctuating networks. The resulting degraded performance or high latency all hinder their widespread applications. 
In this paper, we present Janus, the first framework for low-latency cloud-device collaborative Vision Transformer inference over dynamic networks. Janus overcomes the intrinsic model limitations of ViTs and realizes collaboratively executing ViT models on both cloud and edge devices, achieving low latency, high accuracy, and low communication overhead. Specifically, Janus judiciously combines token pruning techniques with a carefully designed fine-to-coarse model splitting policy and non-static mixed pruning policy. It attains a balance between accuracy and latency by dynamically selecting the optimal pruning level and split point. 
Experimental results across various tasks demonstrate that Janus enhances throughput by up to  5.15\(\times\) and reduces latency violation ratios by up to 98.7\% when compared with baseline approaches under various network environments.
\end{abstract}


\section{Introduction}

The ubiquitous deployment of cameras in various domains, from surveillance to autonomous vehicles \cite{datondji2016survey,guide2023video,klingler2023video}, has led to an exponential increase in the volume of visual data. This data needs to be processed and analyzed with low latency and high accuracy 
to meet the application-level performance needs consistently. 
Vision Transformers (ViTs) have emerged as a powerful alternative to traditional convolutional neural networks (CNNs) in this field, achieving state-of-the-art (SOTA) performances on a variety of computer vision tasks, such as image classification \cite{dosovitskiy2020image}, object detection \cite{li2022exploring}, semantic segmentation \cite{thisanke2023semantic}, and video understanding tasks \cite{wang2023videomaec}.

While ViTs offer unprecedented accuracy, they are computationally expensive, requiring millions of parameters and billions of floating-point operations (FLOPs) \cite{khan2022transformers,han2022survey}, which makes it difficult for real-time analytics. For example, ResNet-50 achieves 80.12\% accuracy in ImageNet-1k classification while ViT-B achieves a higher accuracy 85.43\% but with 8.02$\times$ FLOPs\cite{mmclassification}.

\begin{figure}[t]
\centerline{\includegraphics[width=0.45\textwidth]{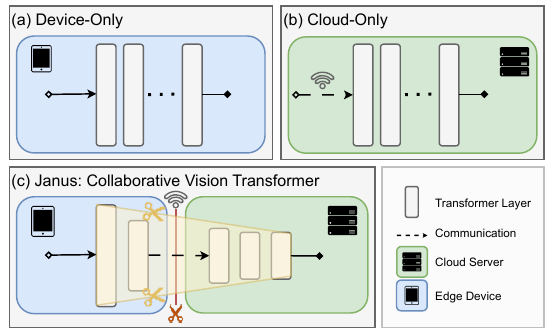} }
\caption{Comparing the existing architectures of serving Vision Transformer (ViT) and Janus: a device-cloud collaborative system that adapts ViT for dynamic networks. }
\label{fig:intro}
\end{figure}
In the deployment of computer vision models, a typical approach is on-device computing (Fig. \ref{fig:intro} (a)), where computational resources on edge devices are often constrained (e.g., the local GPU of Jetson Orin Nano can only serve the ViT-L model for inference at a low-speed of 1.51 FPS). Thus, in this setup, on-device inference often involves the optimization of models. Techniques such as knowledge distillation\cite{hao2022learning}, pruning\cite{zheng2022savit}, quantization\cite{yuan2022ptq4vit}, neural architecture search\cite{you2022shiftaddnas}, and lightweight networks\cite{mehta2021mobilevit} are widely studied to offer competitive service with a smaller model footprint. However, the optimization of models for edge devices still inevitably compromises accuracy and is fundamentally limited by the scarce resources on the device side. 

Another prevailing approach is to perform computer vision tasks in a distant cloud server\cite{zhang2020decomposable}. In this setup (Fig. \ref{fig:intro} (b)), data collected by edge devices are transmitted to a remote cloud server for inference utilizing more powerful accelerators. However, this approach highly depends on network conditions and introduces extra communication delay\cite{laskaridis2020spinna}. 

To address these limitations, a recent line of work\cite{kang2017neurosurgeona, laskaridis2020spinna, yang2022cnnpc} has proposed model splitting as a collaborative approach between device and cloud for low-latency inference. Such schemes typically partition the inference model, often CNNs, between the device and the cloud. At run time, the edge device executes the head part of the model and transmits the intermediate data to a cloud server. The server continues to execute the rest of the model. Overall, this approach decreases data communication latency by transmitting size-reduced intermediate data, thereby achieving low end-to-end (E2E) latency and leveraging the computing power on the cloud side for further computation acceleration.

Despite the significant advantages demonstrated by collaborative inference in serving CNNs, it is non-trivial to apply the same paradigm to serve the emerging ViTs. 
Fundamentally, the key to such an effective collaboration lies in \textit{data reduction} during the inference phase. Namely, by minimizing the volume of data transferred, we can reduce communication latency, ultimately leading to a reduction in E2E latency. While the down-sampling operations inherent in CNN structures naturally reduce input tensor size, minimizing the communication latency of transfer data, a vanilla ViT does not change the input tensor size at all. This naturally deprives ViTs of benefiting from the collaborative framework for further latency reduction.

To push the limit of low-latency ViT inference, in this paper, we introduce Janus\footnote{Janus, a double-faced Roman god overseeing both concrete and abstract dualities, such as life and death, ..., and device and cloud in our case.}, a collaborative cloud-device inference system specifically designed for ViT models that achieves low-latency and high-accuracy inference over dynamic networks (Fig. \ref{fig:intro} (c)). Token pruning in ViTs is an existing technique used to prune redundant image patches and basic input units in ViT models. \textit{Our key insight is that utilizing token pruning to reduce patches can create the opportunity for data reduction and enable potential synergy with model splitting for collaborative inference if carefully designed.}
Janus judiciously integrates the token pruning and the model splitting techniques, supported by carefully designed pruning and splitting policies to realize cloud-device collaboration.
Specifically, the system includes a collaboration-aware token pruner and a fine-to-coarse model splitter. To determine the optimal pruning level and split point, we design a ViT-oriented latency profiler and a dynamic scheduler. 
These components empower Janus to make efficient configuration choices under dynamic network environments, addressing the unique challenges posed by ViTs. To the best of our knowledge, it is the first work that realizes low-latency collaborative ViTs inference. In summary, our contributions are provided as follows:
\begin{itemize}
    \item By analyzing the inference latency and structure of ViTs, we identify the opportunities and challenges of collaborative cloud-device inference for ViTs. 
    \item We introduce Janus, a novel collaborative cloud-device system designed for the low latency inference of the emerging ViTs over dynamic networks. 
    \item We propose a collaboration-aware token pruner that minimizes accuracy degradation. We further design a fine-to-coarse model splitter that reduces the search space and system overheads.
    \item To adapt to dynamic network conditions, we design a profiler and a dynamic scheduler that determine the optimal pruning levels and split points for the token pruner and model splitter.
    \item  We implement and conduct experiments with Janus in real-world devices and network scenarios. Janus exhibits significant improvements over baselines, enhancing throughput by up to 5.15$\times$ and reducing latency violation ratios by up to 98.7\% with minimal accuracy reduction.
\end{itemize}

\section{Background and Motivation}\label{sec:Background and Motivation}

\subsection{Vision Transformers}

ViT\cite{dosovitskiy2020image} is a groundbreaking model series that uses an encoder-only transformer architecture designed for computer vision tasks, without the traditional use of convolution operations.
The core concept of ViTs is treating all the image patches as tokens and constructing multi-head self-attention among them. This self-attention mechanism computes a weighted sum of the input data, where the weights are computed based on the similarity between the patches in the image. This allows the model to discern the significance of different patches, which helps it capture more informative representations when making predictions.

A typical ViT model comprises three key components:

{\bf Embedding}:  In the context of an input image with dimensions $H$ (height), $W$ (width), and $C$ (channels),  the image is first split into $HW/P^2$ patches, where $P$ refers to the patch height and width. Each patch is then flattened to a vector of length $CP^2$ and linearly projected to patch embeddings. Learnable position embeddings are then added to the patch embeddings to retain positional information to have the complete input embedding vector for all patches.

{\bf Transformer Encoder}: The multilayer transformer encoder then transforms input vectors into the same number of output vectors with the same length. The encoder includes a multi-head attention (MHA) layer, followed by a multilayer perceptron (MLP) which provides nonlinearity. The transformer encoder captures long-range dependencies and contextual information in the input data. The output vectors represent the features of the image.

{\bf Heads}: After being processed by the transformer encoder, the output vector is further transformed into the output label through the task-specific neural networks, referred to as heads, to provide predictions for a specific task. For example, ViT usually uses an MLP as the head for image recognition tasks. 

Compared to traditional CNNs, ViT and its subsequent models 
gain advantages in capturing global relationships. Instead of focusing on local features in CNNs, such a global view allows ViT to understand complex visual patterns, making it a SOTA solution in various computer vision tasks\cite{khan2022transformers}.

\subsection{Cloud-only or Device-only: One Size Fits All?} \label{sec:Vision Transformers Latency Measurements}
The rapid inference capabilities of ViTs are especially crucial in low-latency applications.
In the following subsections, we delve into the sources of ViT inference latency, providing insights into the challenges and opportunities for optimizing the inference performance of ViTs. Consider an E2E scenario where users interact with a mobile application leveraging ViTs on their smartphones or vehicle-mounted systems. Throughout the day, users may encounter diverse network conditions, including throughput fluctuations, degradation, and occasional disconnections due to blockage or user equipment mobility. Additionally, they may undergo transitions between different network environments, such as 4G, 5G, or WiFi.
In this pilot study, we evaluate the latency performance of executing ViT on edge devices or cloud platforms under different network situations. 

In the device-only case, our approach involves the execution of ViT on edge devices immediately upon receiving new frames. In the cloud-only case, the edge device transmits the newly captured frame that employs LZW compression\cite{nelson1989lzw}, a commonly used fast lossless compressing algorithm in practice, to the cloud using a specific communication technique—4G, 5G, or WiFi. Subsequently, the cloud processes these frames using the same ViT model. 

{\bf  Measurements Setup: }
For hardware platforms, our edge device platform employs Jetson Orin Nano which features an NVIDIA Ampere GPU with 1024 CUDA cores and 32 Tensor cores, while our cloud platform is equipped with an NVIDIA Tesla V100 GPU with 5120 CUDA cores
and 640 Tensor cores (detailed in Section \ref{sec:Experiment Setup}). For the inference model, we utilize ViT-B \cite{dosovitskiy2020image} as our chosen inference model, representing the initial and most typical model in the family of ViTs.
We use recent reported or measured network statistics for 4G, 5G, and WiFi to estimate the communication latency for our problem. Specifically, the mean upload throughput is 7.6 Mbps and 14.7 Mbps for 4G and 5G, respectively\cite{jackson2023ofcom}. The mean network latency (i.e., round-trip time) is approximately 42.2 ms and 17.05 ms for 4G and 5G measured from a Google Pixel 5 phone to an AWS cloud server\cite{ghoshal2022indeptha}. The mean upload throughput and network latency for a typical 2.4GHz WiFi are 37.68 Mbps and 2.3 ms, according to our measurement.

\begin{figure}[t]
\centerline{\includegraphics[width=0.475\textwidth]{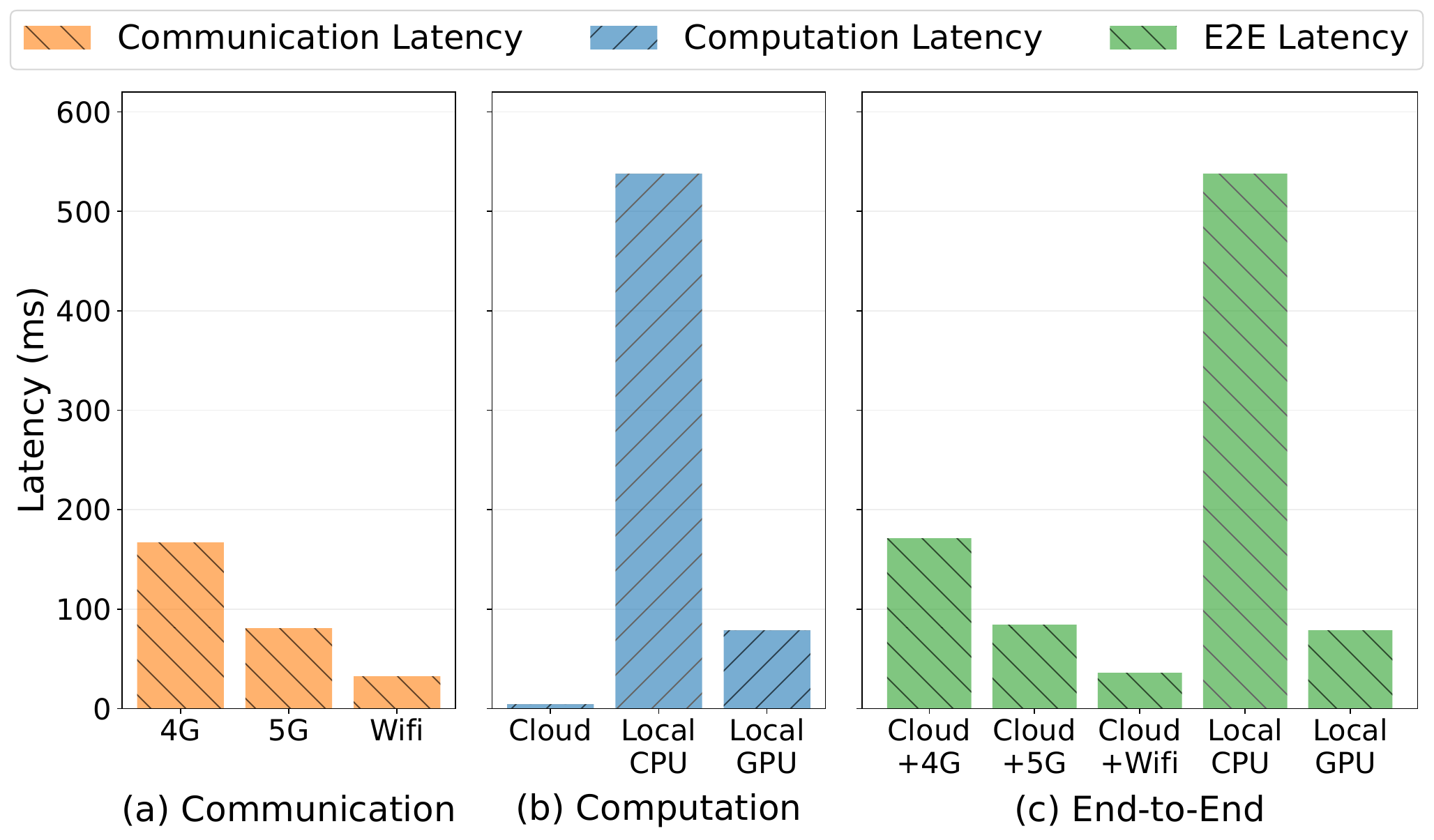}}
\caption{Inference latency breakdown for ViT-B.}
\label{fig:LatencyBreakdown}
\end{figure}

{\bf Latency Breakdown Measurements: }
In Fig. \ref{fig:LatencyBreakdown}, we present a detailed average latency breakdown of a ViT query for an inference on the ImageNet-1k dataset\cite{deng2009imagenet}, where images have a resolution of 224 $\times$ 224. We record the time from sending the frames to the cloud side as communication latency and the execution time running on the device or cloud as computation latency. E2E latency is the sum of computation latency and communication latency required by both cloud and on-device computing. The latency value is the average time required for inference on a single image under these categories. 

Fig. \ref{fig:LatencyBreakdown} (a) shows the communication latency required for uploading a compressed image through 4G, 5G, and WiFi.
Not surprisingly, the 4G and 5G connections have slower performance, requiring 166.84 ms and 80.46 ms, respectively, while WiFi connections exhibit significantly reduced upload times, with 32.17 ms. Fig. \ref{fig:LatencyBreakdown} (b) shows the computation latency on the local CPU, local GPU, and cloud GPU. The local CPU emerges as the slowest platform, demanding 537.42 ms for processing. In contrast, the local GPU and cloud GPU have notably lower latency, with processing times of 78.63 ms and 3.88 ms, respectively.
Fig. \ref{fig:LatencyBreakdown} (c) provides a comprehensive view of the E2E latency required by both cloud and on-device computing under different network situations. In cloud computing, data transfer accounts for a significant portion of the latency, whereas in on-device computing, latency entirely comes from computational processing. In the context of 4G or 5G networks, lower latency is achieved with local GPUs at 78.63 ms, whereas in the context of WiFi networks, better latency is achieved in the cloud at 36.05 ms.

\textit {These observations imply that, given the diversity of available resources on-device and network conditions, cloud-only or device-only inference cannot always deliver an optimal solution for ViTs inference.
This inspires us to explore a collaborative cloud-device solution for robustness and adaptability.}

\subsection{Challenges towards Collaborative ViT Inference} 
Inspired by our measurement insights in Section \ref{sec:Vision Transformers Latency Measurements}, a natural question arises: Is it possible to effectively leverage both the device and cloud resources to enhance the performance of ViTs inference?

For CNN-based vision models, model splitting \cite{kang2017neurosurgeona, matsubara2019distilled, lee2022reconfigurable} is a validated approach to address this challenge. The down-sampling operations in CNNs create opportunities for intermediate data size reduction, consequently making it possible to reduce the data transmission latency during collaboration. 
For instance, when executing AlexNet, a representative model in CNNs, on the ImageNet-1k dataset, the data size after the execution of its Pooling Layer 5 is reduced from 147.88 KB to 26.02 KB, indicating a 95.68\% reduction compared to the input data size of 602.53 KB. The significantly reduced data size suggests a potential benefit of partitioning the CNN model into head and tail models, with the head model (e.g., from the initial layer to the Pooling Layer 5) being executed in the edge device and the tail model (e.g., the remaining layers in the previous example) being executed in the cloud server.
Compared to the cloud-only approach, even though this approach may extend computation time when part of the workload runs on the edge device, it compensates by reducing data transmission and cloud-side computation latency, ultimately resulting in decreased total latency. On the other hand, compared to the device-only approach, despite this approach may incur communication costs when data needs to be transmitted to the cloud, it compensates by reducing device-side computation latency, leading to a decrease in total latency.

Unlike CNNs, vanilla ViTs exhibit distinct structural characteristics. 
In the transformer architecture, the output data size remains relatively consistent. For instance, when considering ViT-B, a typical model in the ViT series, and employing the same input data size as in the previous CNN example, the size of the data after each transformer layer consistently remains at 605.61 KB\footnote{The additional size increase arises from the inclusion of an extra special token introduced by ViT.}. 
\textit{Consequently, existing model-splitting approaches cannot be directly applied to serve transformer architectures due to the absence of substantial data reduction. As a result, when applied to transformers, current model-splitting approaches cannot effectively reduce the communication latency between devices and the cloud.}

\section{System Design}\label{sec:System Design}
This section presents the system design of Janus. To address the challenges revealed by the motivation study, we introduce the first cloud-device collaborative ViT inference framework, named Janus. Janus builds upon the recent development of the token pruning technique and employs a carefully designed dynamic token pruning policy and a model splitting policy to facilitate adaptive and efficient collaborative ViT inference over dynamic networks.

\begin{figure}[t]
\centerline{\includegraphics[width=0.50\textwidth]{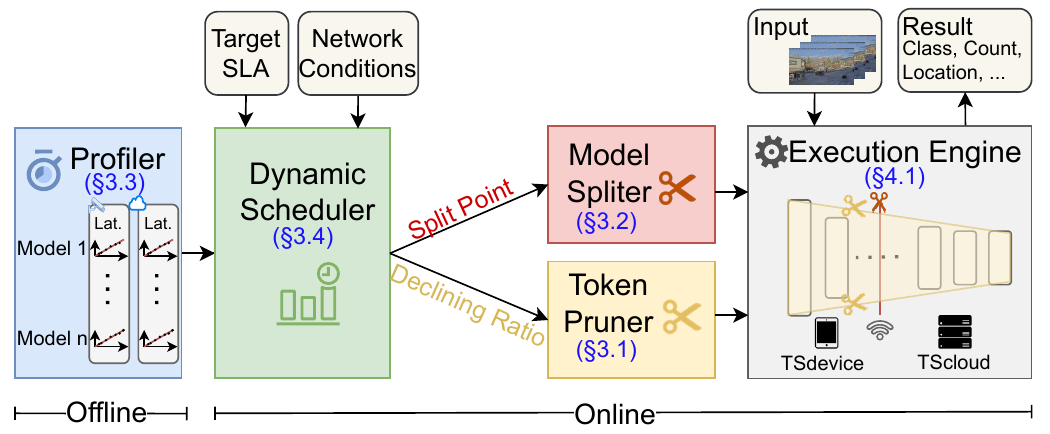} }
    \caption{System overview of Janus.}
\label{fig:Overview}
\end{figure}

Our system, illustrated in Fig. \ref{fig:Overview}, encompasses both offline and online phases. In the offline phase, we deploy a lightweight linear profiler (\S \ref{sec:Profiler}) to predict inference latency under various conditions. At run time, the dynamic scheduler (\S \ref{sec:Dynamic Scheduler}) operates in real-time, leveraging the profiled insights to determine optimal split points and pruning levels based on the target service-level agreement (SLA) for latency and the network environment. Guided by the dynamic scheduler, the collaboration-aware token pruner (\S \ref{sec:Token Pruner}) works with the fine-to-coarse model splitter (\S \ref{sec:Model Splitter}) to effectively prune and split the inference of a ViT model across device and cloud computing environments. 
The execution engine (\S \ref{sec:Execution Engine}) then takes charge of coordinating the inference process and managing communication between partitions.
The details are presented in the following.

\subsection{Collaboration-aware Token Pruner with Mixed Pruning Policy} \label{sec:Token Pruner}
The first module in our system is a collaboration-aware token pruner that realizes dynamic token size reduction under the guidance of the scheduler. 

\textbf{Observation.} Token pruning is a novel model optimization technique in transformer-based models that reduces the number of tokens to be executed\cite{fayyaz2022adaptive, bolya2023token}. Based on the importance or relevance of each token, a subset of tokens is pruned for removal at each transformer layer. While it is mostly studied to accelerate model inference, we observe that it has a huge potential synergy with model splitting to realize collaborative inference due to its data reduction outcome. For instance, we deploy the SOTA pruning approach, ToMe\cite{bolya2023token}, which prunes a fixed number of tokens at each layer. 
Specifically, following the same setting as the original paper, we prune 23 image patches at each layer of the ViT-L@384 in our experiments and find out that 95.7\% of image patches are pruned, demonstrating a pathway to substantial data reduction.

However, how to incorporate it into the collaborative framework so that ViT can be executed in low latency over dynamic networks has not been explored before. 
Sticking to a fixed pruning level, as the current literature does, overlooks the intrinsic characteristics of the underlying computing infrastructures, which leads to suboptimal performance. 
Considering the significant difference in computing capability between on-device and cloud computing, aggressive pruning can significantly reduce computing workload and latency on edge devices. Conversely, given the ample resources in the cloud, excessive pruning in the cloud can lead to reduced accuracy without significant gains in throughput.

\textbf{Design.}
Therefore,  we propose a novel \textit{mixed pruning policy} specifically tailored for cloud-device collaboration. Our policy adopts mixed pruning levels among different layers in ViT. 
We use the term ``declining rate'' to measure the rate at which tokens are pruned in the model during inference and denote it as $\alpha$. For the entire model, a higher declining rate $\alpha$ results in a larger cumulative reduction in tokens, leading to more loss in accuracy. For each layer of the model, the number of pruned tokens decreases with the increasing layer number.

Specifically, we adopt an exponential form to control the extent of token pruning. The number of reduced tokens at each layer $l$ is expressed as follows: 
\begin{equation}
\Delta x_l=\begin{cases}
\lfloor2^{\alpha(N-l)}\rfloor & \text{if } \alpha \neq 0\\
0 & \text{if } \alpha = 0,
\end{cases}
\end{equation}
The declining rate $\alpha$ increases in increments of $t$ within the specified range $\alpha \in \left[0,\alpha_{max} \right]$. When $\alpha = 0$, no pruning occurs. The maximum value for $\alpha_{\text{max}}$ is determined by the following constraint:\\
\begin{equation}
\sum_1^N\lfloor2^{\alpha_{max}(N-(l-1))}\rfloor \leq x_0-1
\end{equation}
where $x_0$ is the initial number of tokens in the model.
This ensures that the cumulative reduction in tokens does not exceed the threshold  $x_0-1$. $N$ is the total number of transformer layers in the ViT model, and $l \in \left[1, N\right]$ denotes the layers in the transformer.

Compared with a linear declining rate where the number of tokens pruned decreases linearly with each layer, $\Delta x_l = \lfloor \alpha \cdot (N - l)\rfloor$, our experiments on ViT-L demonstrate that the exponential-form declining rate reduces latency more, particularly on edge devices, with almost the same accuracy loss (less than 0.0021). More statistics are in Table \ref{tab:pruning poliy}. 

\begin{table}[t]
\centering
\caption{Comparison of latency reduction with different pruning strategies on edge device and cloud server.}
\label{tab:pruning poliy}
\begin{tabular}{c|c|c}
\hline 
\multirow{2}{*}{\textbf{Pruning Strategy}} & \multicolumn{2}{c}{\textbf{Latency (ms)}} \\ \cline{2-3} 
& \textbf{Edge Device} & \textbf{Cloud Server} \\ \hline 
No Pruning & 653.3 & 32.3 \\ \hline 
Linear Declining & 432.0(-221.3) & 24.2 (-8.0) \\ \hline 
Exponential Declining & 403.2(-250.1) & 22.5 (-9.8) \\ \hline 
\end{tabular}
\end{table}

\subsection{Fine-to-Coarse Splitting Points Generation} \label{sec:Model Splitter}
After pruning, we design a model splitter to partition execution between a device and a server.
As a model consisting of a sequence of layers, it allows us to split the model at the granularity of individual layers. For a ViT with $N$ transformer layers, there are $N+2$ candidate split points within the model, including the point at the very beginning of the model, after each of the N transformer layers, and at the end of the model.

\textbf{Observation.} As we mentioned before, the decrease in latency primarily comes from the reduction in transferred data. In the declining pruning policy, the front part experiences more data reduction, leading to a greater reduction in data transmission latency, which makes it easier to identify the highest-performing split point that brings the lowest latency. Conversely, the rear part experiences less data reduction, resulting in a smaller decrease in data transmission latency. Even if a split point with the lowest latency is identified in the rear part, its latency difference with the surrounding candidate split points is minimal during model inference. For example, in the ViT-L model with a pruning ratio $\alpha$ set at 0.25, the front part experiences approximately 90\% data reduction, while the rear part contributes only 10\% to the total data reduction. As the split point moves from the front to the rear, the benefits in transmission latency reduction gradually diminish. Based on this key observation, we decide to focus more on layers in the front portion where the potential for latency reduction is more significant, instead of considering splitting at each layer uniformly. This design allows us to reduce the search space of the partitioning points and overall system overheads.

\textbf{Design.}
Consequently, we design a \textit{fine-to-coarse splitting points generating policy}, aligning with the declining mixed pruning policy. 
In our fine-to-coarse splitting policy, more candidate points are set in the front part of the model, while fewer are in the rear part.  
Formally, the candidate split point set, denoted as $C$,  is defined as the union of two sets: 
\begin{equation}
\begin{split}
C = \{0, N+1\} \cup \{ s_i \mid s_i = s_{i-1} + \left\lceil \frac{i}{k} \right\rceil, \\ s_1 = 1, i \geq 2 \text{ and } s_i \leq N \}
\end{split}
\end{equation}

\begin{figure}[t]
\centerline{\includegraphics[width=0.5\textwidth]{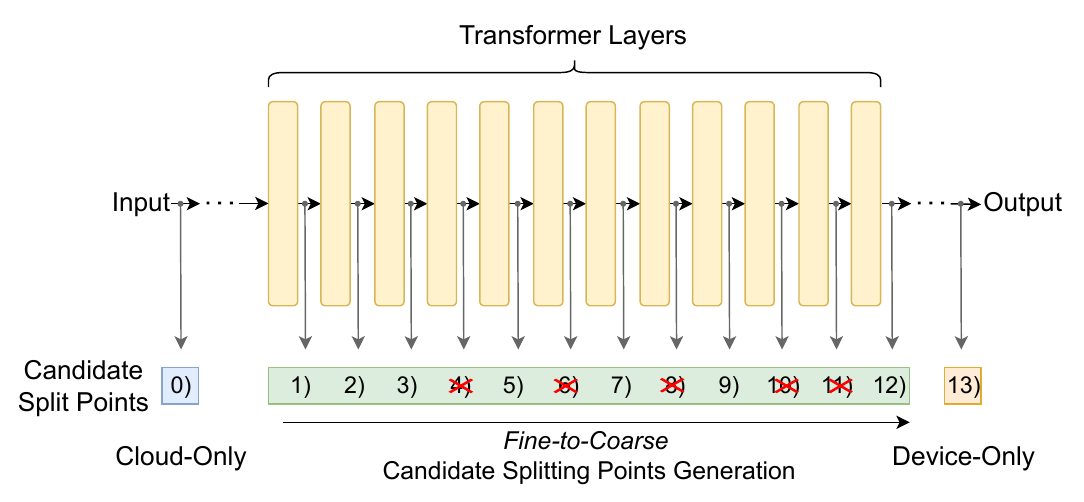} }
\caption{The fine-to-coarse candidate splitting points generating policy applied to a ViT with 12 layers when k is set to 3. The crossed indexes represent split points that are removed after applying the splitting policy.}
\label{fig:SplitPolicy}
\end{figure}

For the first set, it indicates that the endpoints $s = 0$ and $s = N+1$ (indicating cloud-only and device-only processing, respectively) are always considered candidate split points. For the second set, concerning split points $s$ within the range $\left[1, N \right]$ (indicating splits after each transformer layer), the process begins with $s_1=1$ and continues determining candidate split points until $s_i$ exceeds $N$. Here, $i$ is the index of candidate split points, and the parameter $k$ controls their quantity. $k$ regulates the spacing between candidate split points; a smaller $k$ value leads to a denser distribution of candidate split points, resulting in more candidate split points. Fig. \ref{fig:SplitPolicy} shows an example using a ViT with 12 layers when k is set to 3. The crossed indexes represent split points that are not considered after applying our split policy. This approach helps reduce the search space of the candidate split points, thereby reducing the execution time of the dynamic scheduler (Section \ref{sec:Dynamic Scheduler}) and consequently minimizing the overall system overheads.

\subsection{Lightweight Linear Profiler}  \label{sec:Profiler}
Given the varying trade-offs of different split points and declining rates, we need to consider latency constraints, network conditions, and expected latency to select the most suitable configuration. While the latency constraint is given by the user and the network condition estimation has been widely studied, it remains unknown how to determine the 
expected computation latency for each candidate point. 
\begin{figure}[t]
\centerline{\includegraphics[width=0.5\textwidth]{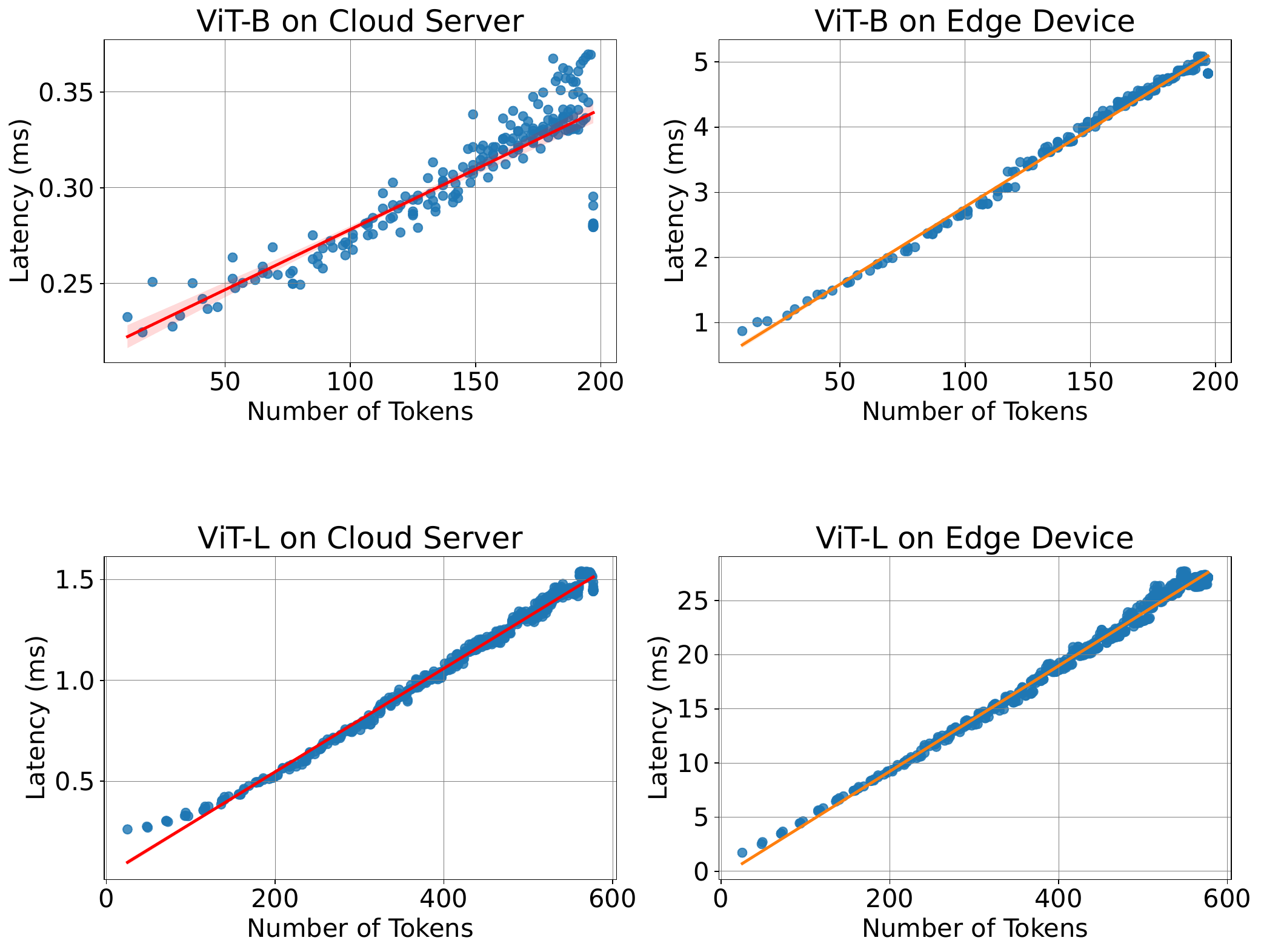} }
\caption{Layer latency of ViTs across different numbers of tokens.}
\label{fig:Lat2toknum}
\end{figure}

\textbf{Observation.}
We first conduct an experiment involving random pruning of ViT layers, observing the latency of each transformer layer. Experiments use different ViT models, including ViT-B and ViT-L$@$384(more detailed setup in Section \ref{sec:Experiment Setup}). Fig.  \ref{fig:Lat2toknum} illustrates the experimental results, showing the average layer inference latency for different numbers of input tokens per layer. As can be seen, for either the edge device or the cloud server, the inference latency of each layer exhibits a strong positive linear relationship with the number of its input tokens. Notably, in all cases, the correlation coefficient exceeds 0.85 and the P-value is very close to 0.

\textbf{Design.} Based on these observations, we propose a simple but effective linear model for each ViT as a profiler. The profiler adopts a linear function as the prediction model, which takes the number of input tokens per layer as input and outputs the predicted inference latency. We use linear regression to calculate the coefficient for this model. This choice aligns with real-world behaviors and offers the advantage of reducing computational overhead for the profiler, simplifying the generation of prediction models.

\subsection{Dynamic Scheduler} \label{sec:Dynamic Scheduler}

Existing token pruning techniques typically prune a fixed number of tokens without considering the variability of network conditions. 
To address this, we design a dynamic scheduler that identifies the highest-performing split point and pruning level in dynamic network environments.

The scheduling algorithm, outlined in Algorithm \ref{algorithm:dynamic scheduler}, first explores a range of declining rates $\alpha$ from 0 to $\alpha_{\text{max}}$, traversing the model's accuracy from high to low (lines 2-4). Once a declining rate is chosen, the algorithm computes the number of tokens $x_l$ at each layer based on the selected ratio.
Using prediction models in the profiler, the algorithm predicts the device latency $T^{device}_l$ for $x_l$ tokens running on the device and the cloud latency $T^{cloud}_l$ for the same layer running on the server (lines 6-7). Additionally, the communication latency $T^{comm}_l$ is calculated under the estimated current bandwidth $B$, given the input data size (line 9). 
We estimate the bandwidth based on the harmonic mean of the observed throughput\footnote{Other advanced bandwidth prediction methods can also be applied.}\cite{jiang2012improving}. 
During the cold-start period, we use the mean value of network bandwidth at the offline phase as a rough estimation.
The algorithm then determines the split point $\mathrm{L}_{\mathrm{s}, \alpha}$ that minimizes the overall latency (line 12).
If the identified split point meets the latency requirement $SLA$, the algorithm returns the declining rate $\alpha$ and split point $s$ (lines 13-15). This algorithm finds the configuration with maximum accuracy while satisfying latency requirements. If no combination fulfills the latency requirement, the algorithm returns the maximum declining rate $\alpha_{\text{max}}$ and the split point $s$ that offers the lowest latency among all configurations (line 17).

\begin{algorithm}[t] 
\DontPrintSemicolon
  \KwInput
  {
    \\
    Specific ViT model $M$; 
    Number of layers in the ViT $N$; 
    Layer in the ViT $\{l \mid l=1 \cdots N\}$; 
    Prediction model $f^M\left(X_l\right)$ that returns the latency of executing $X_l$ tokens; 
    Estimated current bandwidth $B$; 
    Data size of each token $D_M$; 
    Latency requirement $SLA$
  }
  \KwOutput
  {
    \\
    Selection of declining rate and split point
  }
  {\bf Procedure}\\
   \For{$\alpha \leftarrow 0$ \KwTo $\alpha_{max}$}
   {
    Choose $\alpha$ as the declining rate\\
    $ \rightarrow  $ $\left\{x_l \mid l=1 \cdots N\right\}$: number of tokens at layer $l$\\
    \For{$\l \leftarrow 0$ \KwTo $N+1$}
    {
        $T^{device}_l \leftarrow f_{\text {device }}^M\left(x_l\right)$\\
        $T^{cloud}_l \leftarrow f_{\text {cloud }}^M\left(x_l\right)$\\
        \If{$l \in C$}{
            $T^{comm}_l \leftarrow \frac{x_l * D_M}{B}$\\
        }
    }
    $\mathrm{L}_{\mathrm{s}, \alpha}=\mathop{\arg\min}\limits_{s \in C}\left(\sum_{l=1}^{s} T^{device}_l+\sum_{l=s+1}^{\mathrm{N}} T^{cloud}_l+T^{comm}_s\right)$\\
    \If{$\mathrm{L}_{\mathrm{s}, \alpha} \leq$ $SLA$ }
    {
        {\bf return}  declining rate $\alpha$ and split point $s$
    }
   }
   {\bf return} declining rate $\alpha_{max}$ and split point $s$ \textcolor{blue}{\scriptsize{\tcp{cannot meet the latency requirement}}}
\caption{Workflow of dynamic scheduler}
\label{algorithm:dynamic scheduler}
\end{algorithm}

The overall time complexity of the scheduling algorithm is $O(\frac{\alpha_{max}}{t} \times N)$. Experimental observations indicate that the average execution time is around 1 ms, which shows the small overhead our scheduler introduces.

\section{Janus Runtime}\label{sec:Runtime}
This section describes the Janus runtime and the implementation of the prototype.
\subsection{Execution Engine} \label{sec:Execution Engine}
We develop two customized modules as execution components of the Janus system. These modules are the \textbf{Jdevice}, which serves as the edge device infrastructure, and the \textbf{Jcloud}, which serves as the cloud server infrastructure. 

In {\bf Jdevice}, both the profiler and the dynamic scheduler are deployed on the device side. When an inference task arrives, the system reads the model type and latency requirements. The profiler gives corresponding latency prediction models and estimates the network conditions. Subsequently, the dynamic scheduler determines deployment parameters, including the split point and declining rate. The deployment parameters, along with the model type, are then transmitted to the Jcloud. Based on deployment parameters, the customized device-side ViT model is then prepared and executed. Its intermediate output is compressed using standard LZW compression before transmission to the Jcloud.

In {\bf Jcloud}, it initially receives information about the model type, split point, and declining rate and prepares the customized cloud-side ViT model. It then receives and decompresses the intermediate output from the Jdevice. Using this intermediate result as input, the cloud-side model runs to obtain the final inference results.

\subsection{Implementation}
The implementation of Janus comprises about 2.53K lines of code (LoC), distributed as follows: 39\% for ViT model operation modules, including the model splitter and token pruner; 27\% for the profiler and dynamic scheduler; and 34\% for communication between the Jdevice and Jcloud.
Modules in Janus are all implemented in Python for easy compatibility with ViTs. Communication of compressed intermediate data, control messages, and prediction results between the edge device and the cloud server is passed via general socket APIs. Image and Video processing is implemented using the OpenCV library\cite{opencv}. Model operations, including model splitting and token pruning, are based on the timm (PyTorch Image Models) library\cite{pytorch}.
\begin{figure}[t]
\centerline{\includegraphics[width=0.45\textwidth]{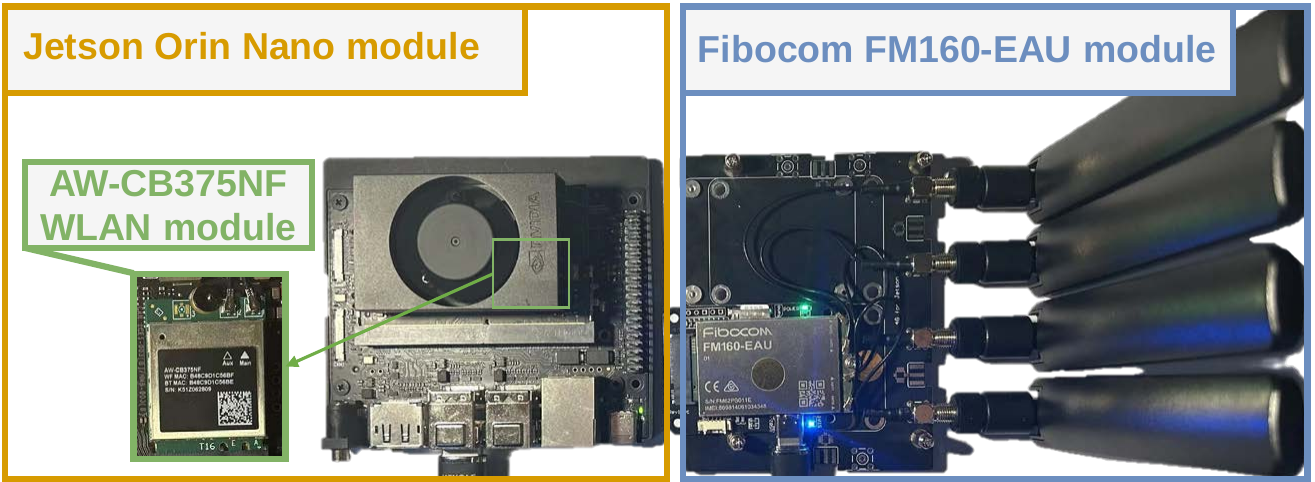} }
\caption{Devices in our real-world deployment.}
\label{fig:Deployment}
\end{figure}
\begin{figure*}[t]
    \centering
    \subfloat[\label{subfig:img_cls_lte} Image recognition on LTE.]{
        \includegraphics[width=1.70in]{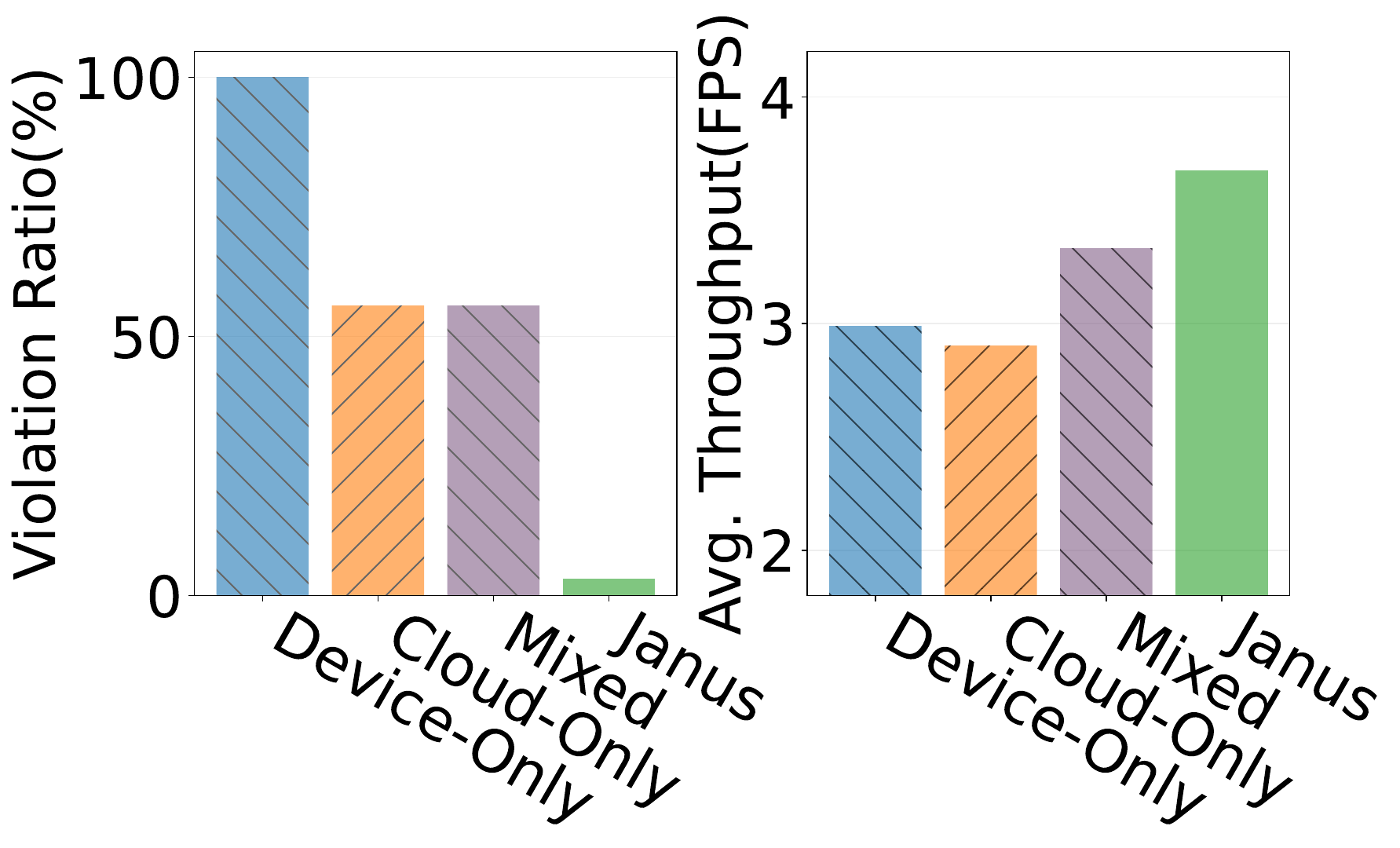}
    }
    \subfloat[\label{subfig:img_cls_5g} Image recognition on 5G.]{
        \includegraphics[width=1.70in]{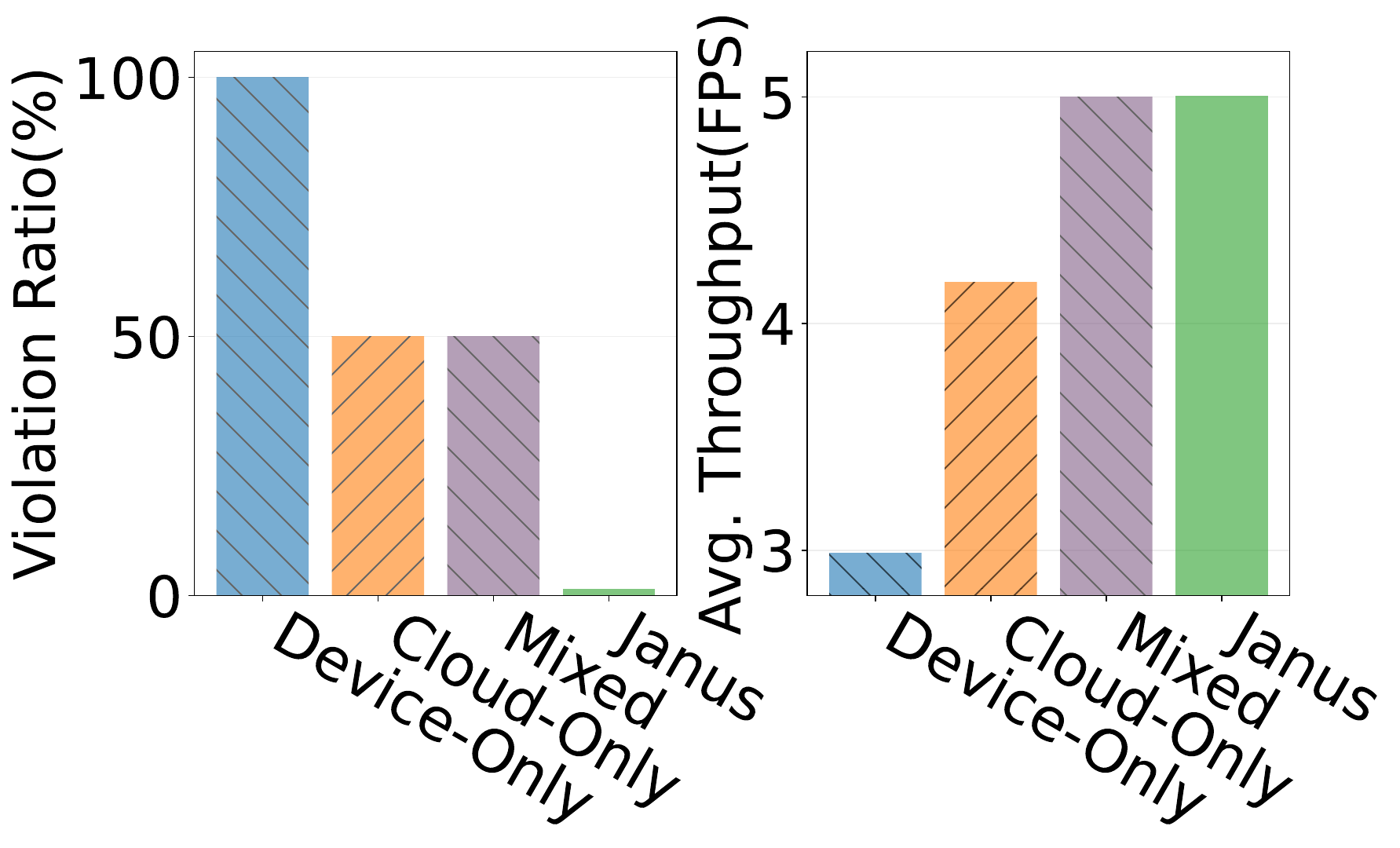}
    }
    \subfloat[\label{subfig:vid_cls_lte} Video classification on LTE.]{
        \includegraphics[width=1.70in]{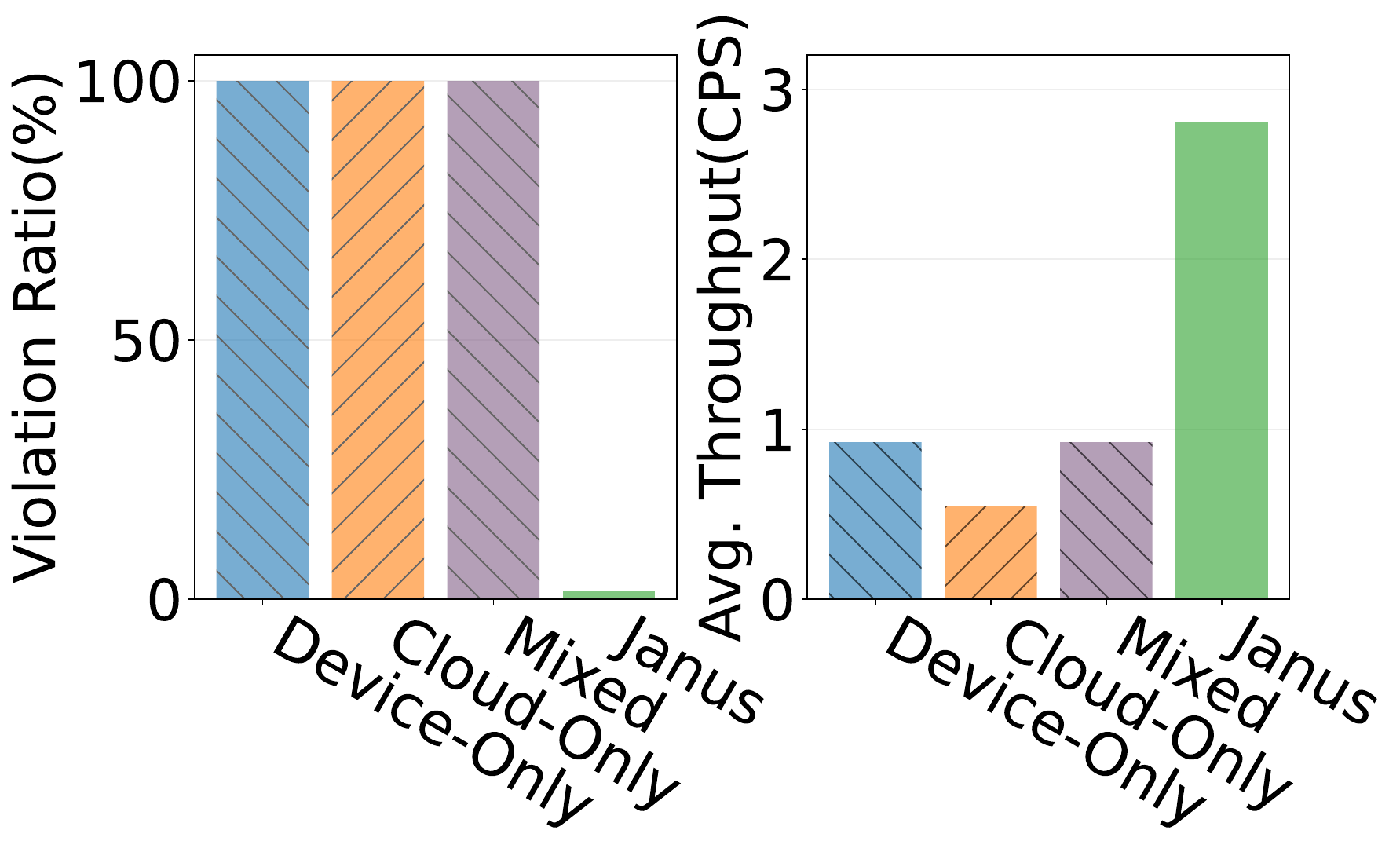}
    }
    \subfloat[\label{subfig:vid_cls_5g} Video classification on 5G.]{
        \includegraphics[width=1.70in]{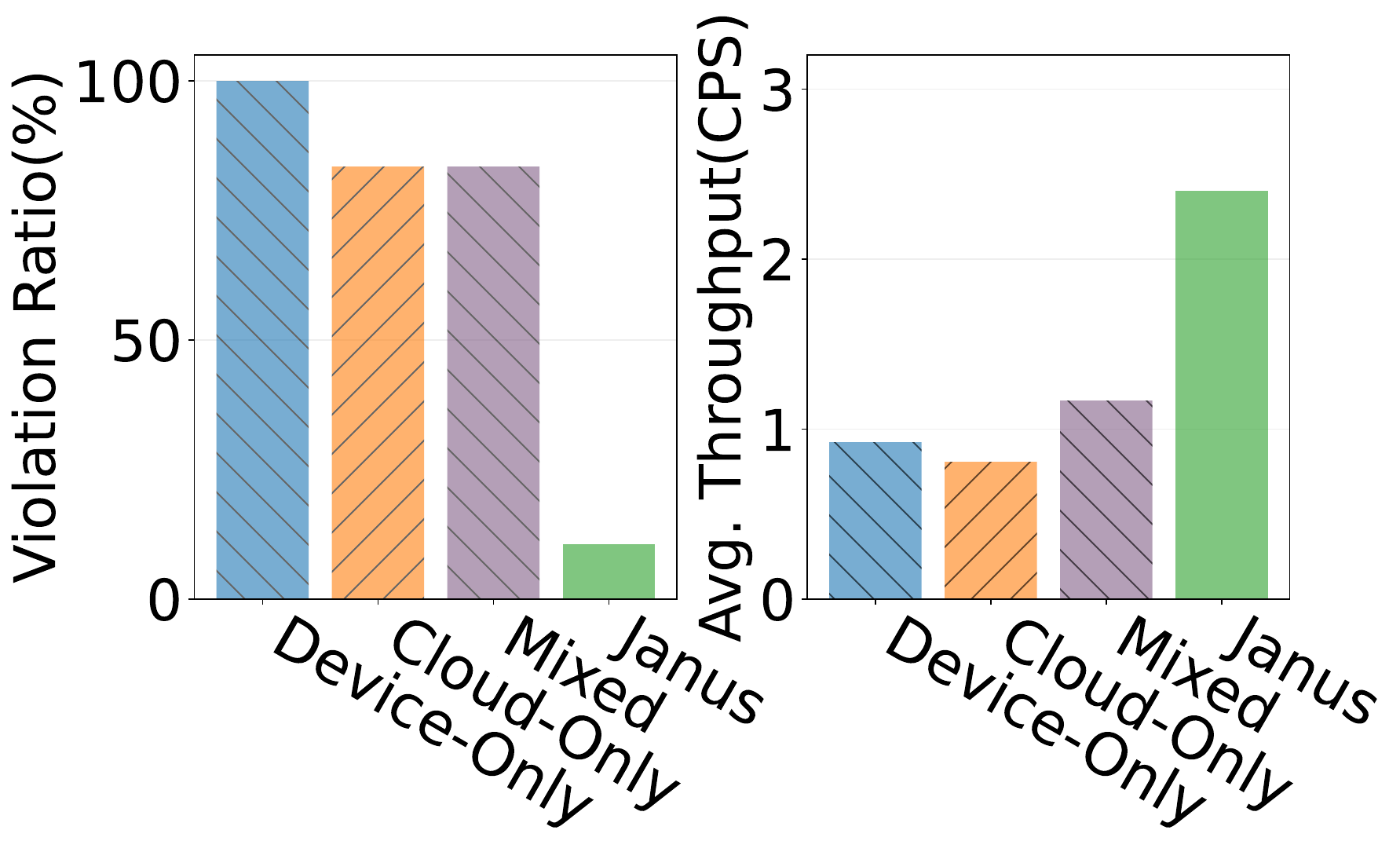}
    }
    \caption{The overall performance under different network conditions and tasks. Janus shows a slight improvement in average accuracy compared to the baselines.}
    \label{fig:gen-res}
\end{figure*}
\section{Evaluation}\label{sec:Evaluation}
Our evaluation aims to answer three main questions: 
\begin{itemize}
    \item Does Janus achieve low latency while maintaining accuracy under various dynamic network conditions?
    \item How do network conditions affect the performance of Janus?
    \item How much overhead does Janus introduce? 
\end{itemize}

We answer the first question by performing a simulation experiment on real-world dynamic network trace on different ViTs inference tasks (\S \ref{sec:Janus Performance}); the second by conducting a sensitivity analysis to network conditions (\S \ref{sec:Sensitivity Analysis}); and the third by conducting an overhead analysis using real-world deployment (\S \ref{sec:Overhead Analysis}). Before presenting our results, we first detail the real-world deployment of our prototype and experimental setup as follows.

\subsection{Real-World Deployment} \label{sec:Deployment}
We deploy the prototype of Janus with a commodity device and a public cloud computing platform. As shown in Fig. \ref{fig:Deployment}, the edge device is an NVIDIA Jetson Orin Nano\cite{nvidia}, a widely used device computing platform for existing video analytics systems. It is equipped with an Ampere GPU with 1024 CUDA cores and 32 Tensor cores, an ARM Cortex-A78AE CPU, and 8GB LPDDR5 DRAM. The edge device uses an AW-CB375NF wireless network module (for WiFi communication) or a Fibocom FM160-EAU module (for cellular mobile communication, such as 4G or 5G) to transmit data to a server. The cloud server is an Aliyun ECS Instance\cite{elastic}, with 4$\times$Intel(R) Xeon Platinum 8163 Processor @ 2.50GHz, 32GB RAM, and Tesla V100 SXM2 with 5120 CUDA cores and 640 Tensor cores, representing a typical cloud server. Both the edge device and the cloud server operated on Ubuntu 20.04 LTS OS and utilized PyTorch 2.0.0. \\

\subsection{Experiment Setup} \label{sec:Experiment Setup}
{\bf Vision datasets and tasks:} We evaluate two typical computer vision tasks: 1) A frame-level image recognition task on the ImageNet-1k dataset \cite{deng2009imagenet}. 
2) A video-level video classification task on Kinetics-400 \cite{kay2017kinetics}. Videos are stored on the edge device and fed into the system video clip by video clip. We adopt the inference protocol of 10 clips $\times$ 1 crop on Kinetics-400 for the evaluation.\\
{\bf Network trace datasets:} To assess the effectiveness of Janus under real-world network conditions, we utilize the 5G mmWave uplink performance dataset \cite{ghoshal2022indeptha} for the simulation experiment. The dataset includes three measurement scenarios: Static, Walking, and Driving, covering 5G and 4G LTE networks. We utilize this dataset for its dynamic network conditions, which include high levels of fluctuations due to challenging scenarios such as blockage and mobility.\\
{\bf Parameter settings:}
For the token pruner, we set the value of parameter $t$ to 0.01. For the model splitter, we set the value of parameter $k$ to 5. For the frame-level \textit{image recognition task}, the model we use is ViT-L@384 \cite{dosovitskiy2020image}. The input frame size is 3$\times$384$\times$384 and the patch size is 16$\times$16. The number of layers in the ViT $N$ is 24. The latency requirement $SLA$ is set to 300 ms per frame.
For the video-level \textit{video classification task}, the model we use is ViT-L from Spatiotemporal MAE\cite{feichtenhofer2022masked}. The input clip size is 16$\times$224$\times$224 and the patch size is 2$\times$16$\times$16. The number of layers in the ViT $N$ is 24. The latency requirement $SLA$ is set to 600 ms per clip. The $SLA$ values are used by default unless we study the impacts of various latency requirements.\\
{\bf Baselines:} We compare Janus with the following baselines:
1) \textit{Device-Only}: The processing is only conducted on the edge device.
2) \textit{Cloud-Only}: All computations are offloaded to the cloud, with local data compressed by the LZW algorithm and transmitted to the cloud server.
3) \textit{Mixed}: According to the profiler, this approach selects between Device-Only and Cloud-Only to minimize the predicted latency based on the dynamic network conditions. Notably, NeuroSurgeon\cite{kang2017neurosurgeona}, a classical method in cloud-edge collaboration for serving CNNs, transitions to the Mixed approach when serving ViTs.
For each of these baselines, we applied the maximum fixed token pruning levels as reported by the SOTA method ToMe \cite{bolya2023token}: pruning 23 tokens per layer for the ViT-L@384 and 65 tokens per layer for the ViT-L from Spatiotemporal MAE.\\
{\bf Evaluation metrics: }We assess the performance of Janus and the baselines along the following evaluation metrics: 
1) \textit{Latency Requirement Violation Ratio}. It is the percentage of inferences that did not meet the latency constraint.
2) \textit{Inference Accuracy}. It is the percentage of frames/clips correctly classified.
3) \textit{Average Throughput}. It is the average frames/clips being inferred per second (FPS/CPS).
4) \textit{Latency Deviation Rate}. It is the percentage difference between measured latency and required latency($\max\left(0, \frac{{\text{Latency}_{\text{measured}} - \text{Latency}_{\text{requirement}}}}{{\text{Latency}_{\text{requirement}}}}\right)$).

\subsection{Janus Performance} \label{sec:Janus Performance}

{\bf Overall Improvements:}
We first present the overall performance of Janus and baselines on different network conditions and tasks in Fig. \ref{fig:gen-res}.  
Janus successfully addresses the shortcomings of the baselines, achieving the highest performance across diverse network conditions for both tasks. 
Specifically, compared to Device-only, Cloud-only, and Mixed approaches, Janus improves average throughput by a factor ranging from 1.23$\times$ to 3.04$\times$, 1.20$\times$ to 5.15$\times$ and 1.00$\times$ to 3.04$\times$, respectively, while reducing violation ratios by a range of 89.4\% to 98.7\%, 49.8\% to 98.3\% and 49.8\% to 98.3\%, respectively. 
Janus achieves an average accuracy improvement ranging from 0.01\% to 0.29\% over the baselines, demonstrating that it provides higher throughput with a smaller reduction in accuracy compared to the baseline methods.

{\bf Understanding Janus Improvements:}
To understand the reasons behind Janus yielding such benefits, we present a representative period extracted from the bandwidth trace for further illustrations. For brevity, we focus on the image recognition task under LTE network conditions.

\begin{figure}[t]
\centerline{\includegraphics[width=0.5\textwidth]{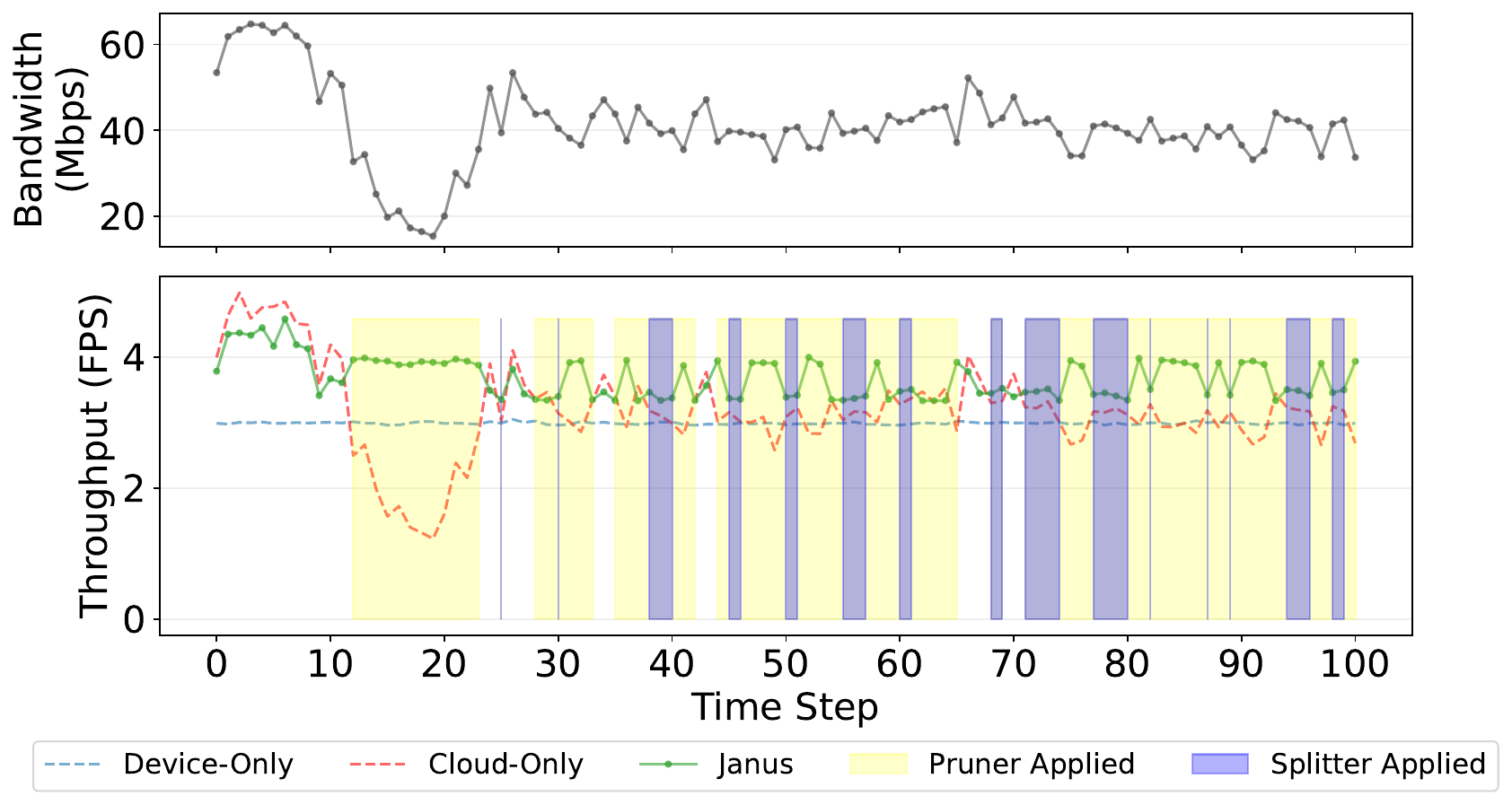} }
\caption{Illustration of how Janus works for the image recognition task in a 4G LTE network trace.}
\label{fig:reason}
\end{figure}

\begin{figure*}[t]
	\centering
	\subfloat[\label{a} Latency of image recognition task]{
		\includegraphics[width=1.73in]{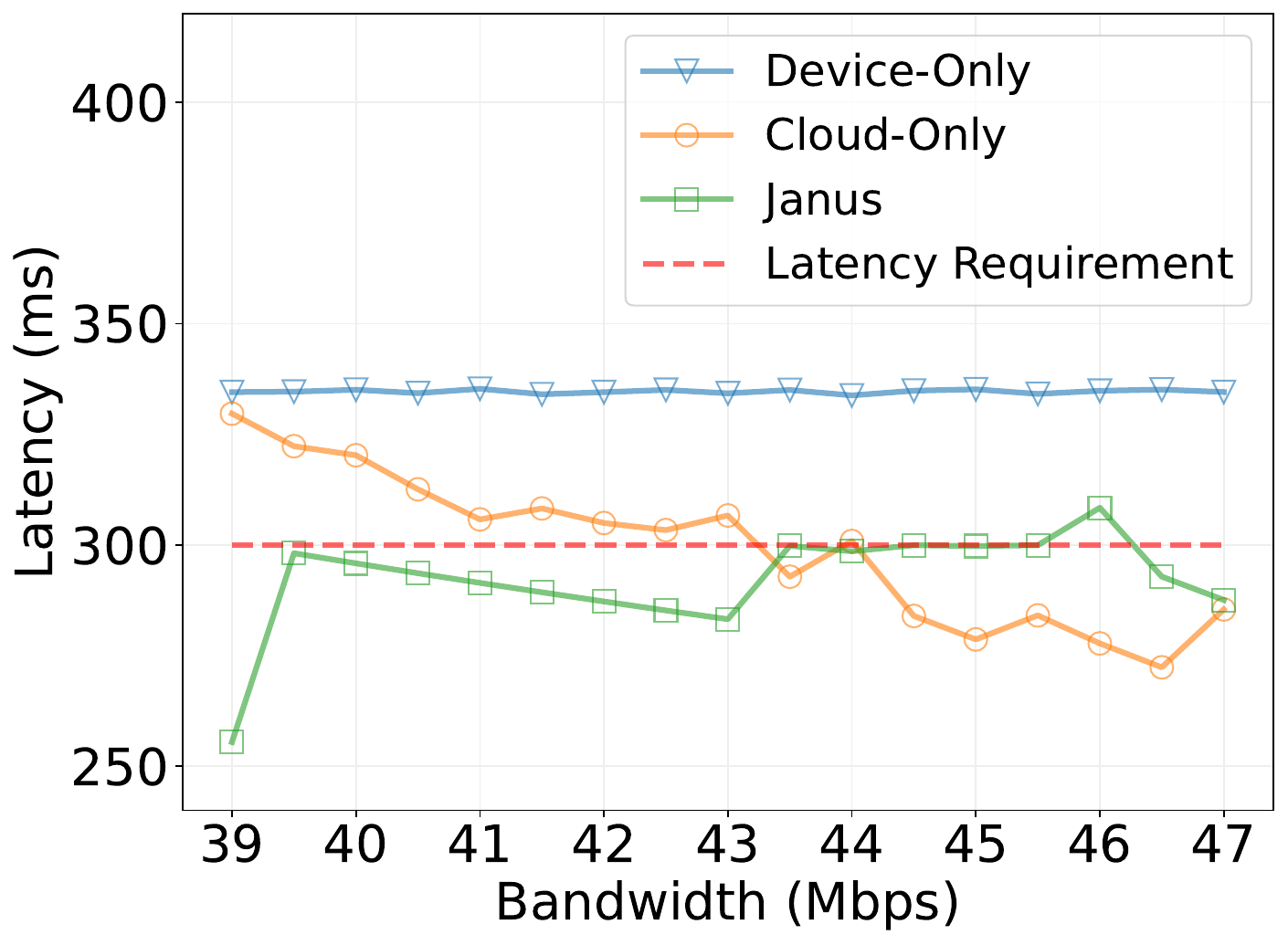}}
	\subfloat[\label{b} Selection of image recognition task]{
		\includegraphics[width=1.73in]{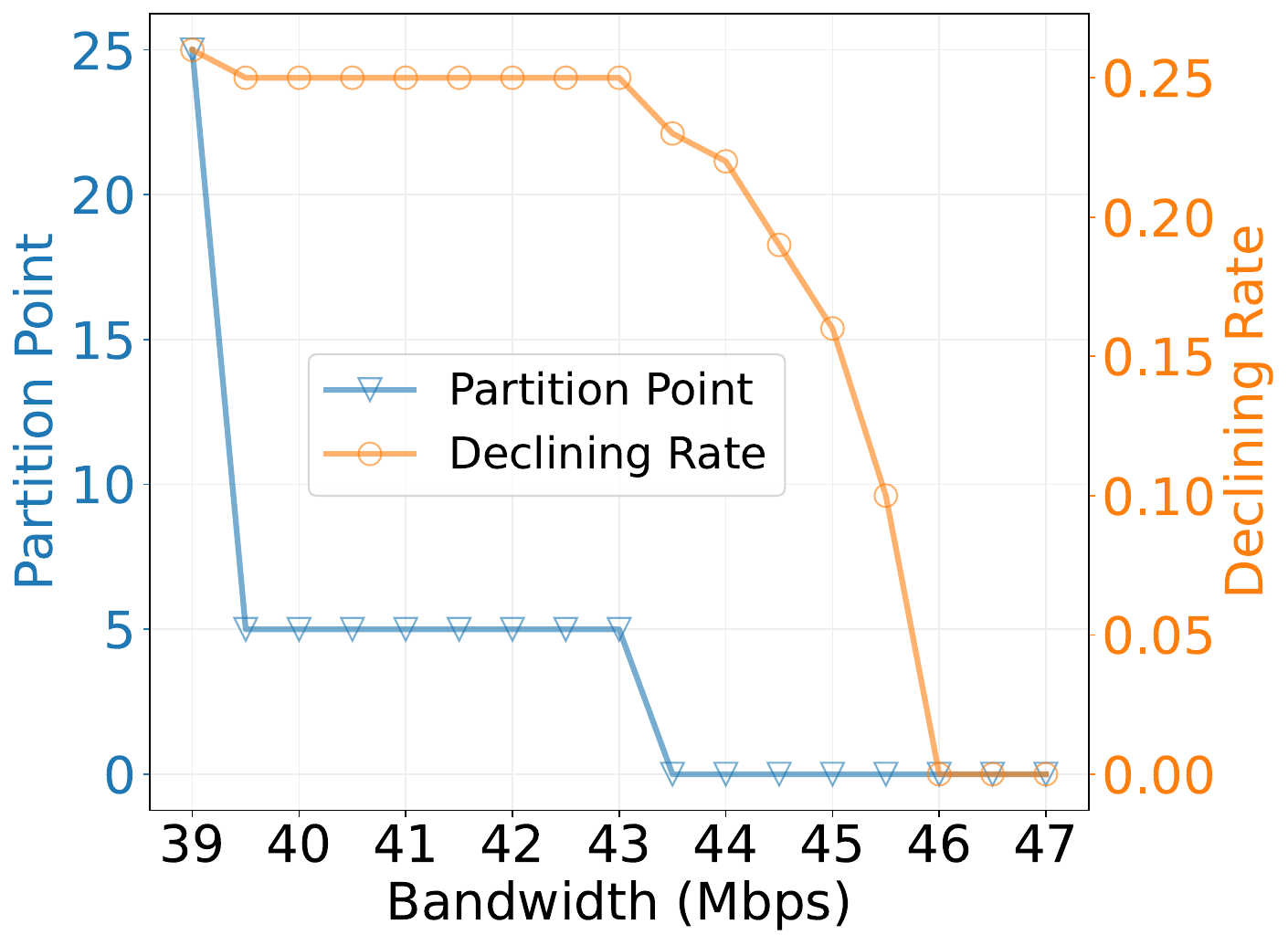} }
    \subfloat[\label{a} Latency of video classification task]{
		\includegraphics[width=1.73in]{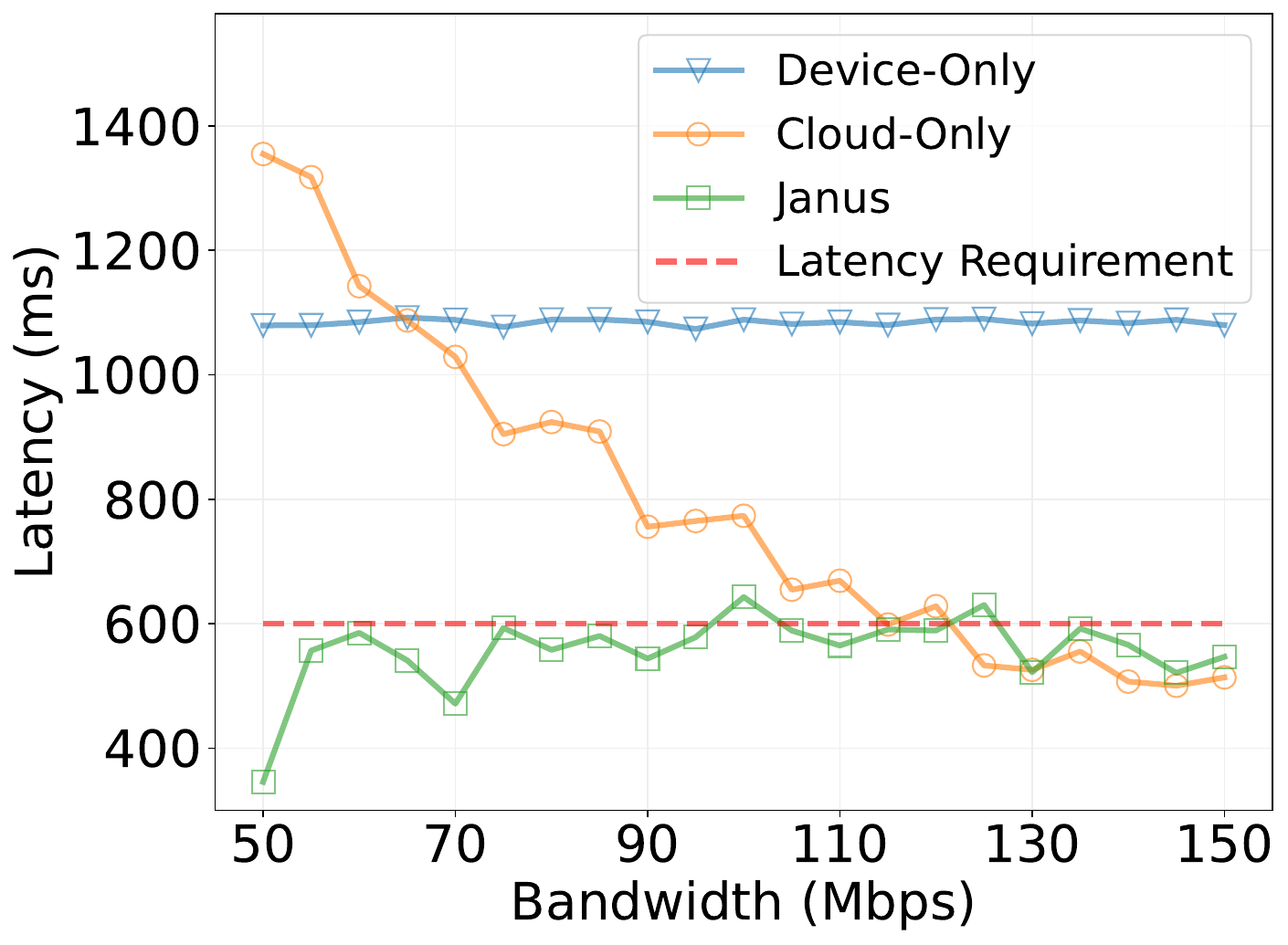}}
	\subfloat[\label{b} Selection of video classification task]{
		\includegraphics[width=1.73in]{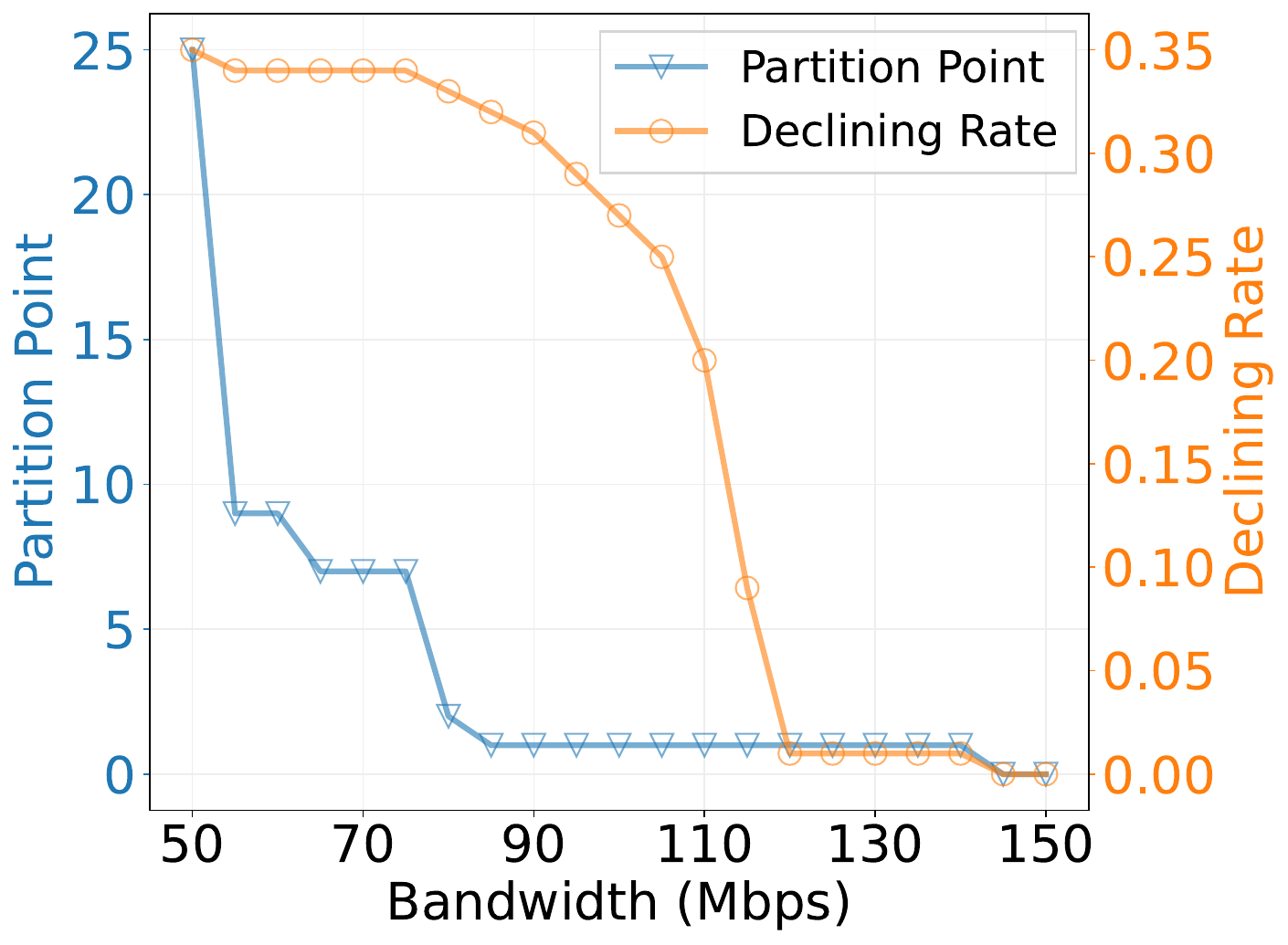} }
	\caption{Latency and the corresponding decision results under varying bandwidths.}
	\label{fig:var-bandwidth}
\end{figure*}
In Fig. \ref{fig:reason}, the top figure illustrates bandwidth traces of 4G LTE (LTE Driving, Run 8 trace) networks over a specific duration. The bottom figure depicts the throughput of Cloud-Only, Device-Only, and Janus in the corresponding network environments. Yellow-shaded regions represent the application of model splitting to partition the model (excluding Cloud-Only or Device-Only inference), while blue-shaded regions indicate the application of token pruning to reduce tokens. Notably, due to the necessity of token pruning occurring before model splitting, the blue-shaded regions cover up the yellow-shaded regions in the figure.

When network conditions are better (e.g., time step $<$ 12), the communication latency is no longer the primary bottleneck in the system. Janus chooses to offload all computing to the cloud without any pruning(same as Cloud-Only processing), as the Cloud-Only approach is sufficient to meet latency requirements. 
This is also the reason why Janus improves less under 5G than 4G network conditions as shown in Fig. \ref{fig:gen-res}. 
In scenarios with relatively low bandwidth(e.g., 12 $\leq$ time step $<$ 24), Cloud-Only suffers significant throughput degradation, and Device-Only maintains low throughput for executing heavy models.
In contrast, Janus shows more stable throughput across various network conditions by collaborative device-cloud inference. 
In scenarios with a relatively high bandwidth(e.g., time step $\geq$ 23), the communication latency decreases. Janus uses the saved latency to maintain accuracy by employing a smaller pruning level while appropriately splitting the model between the cloud and the device.

In summary, Janus utilizes the dynamic scheduler to incorporate model splitting and token pruning, demonstrating adaptability to dynamic network conditions and consistently providing high throughput with minimal accuracy cost.

\subsection{Sensitivity Analysis} \label{sec:Sensitivity Analysis}
{\bf Sensitivity to Network Settings:}
We investigate the impact of varying bandwidth on both tasks.
Fig. \ref{fig:var-bandwidth} (a) and Fig.  \ref{fig:var-bandwidth} (c) show the variation in inference latency of different tasks under increasing bandwidth. Fig.  \ref{fig:var-bandwidth} (b) and Fig.  \ref{fig:var-bandwidth} (d) offer insights into the declining rate and split points selected by the dynamic scheduler in these network conditions. Janus consistently meets the latency requirement in almost all cases, demonstrating its effectiveness. In contrast, Cloud-Only processing can finally meet latency requirements with improved network conditions (e.g., $>$ 44 Mbps in image recognition task).
Underneath, as bandwidth increases, both the declining rate and split point become smaller. The dynamic scheduler progressively offloads more computational tasks from the device to the cloud. Simultaneously, there is a reduction in pruning intensity, resulting in improved inference accuracy. This indicates that Janus can adapt to variations in network conditions to balance accuracy and latency.

\subsection{Overhead Analysis} \label{sec:Overhead Analysis}
We conducted an overhead analysis based on the prototype. The E2E latency of Janus mainly comes from four modules, including system overhead, data transmission, device computing, and cloud computing. The breakdown of their time consumption in real-world scenarios is illustrated in TABLE \ref{tab:overhead}. Due to limited space, we only present the detailed results for the image recognition task. 
In real-world deployments, we tested Janus across WiFi, 5G, and 4G networks, with average bandwidths of 29.3 Mbps, 17.8 Mbps, and 10.1 Mbps, respectively. 
To accommodate the network conditions, we set latency constraints of 500ms for image recognition tasks and 1500ms for video classification tasks. 
As can be seen, as network conditions degrade, the system overhead of Janus increases. This is primarily due to the dynamic scheduler needing to search for larger declining rate values to ensure the task meets latency constraints. The overall system overhead is less than 0.21\%, indicating Janus is lightweight.

\begin{table}[t]
    \centering
    \caption{The overhead analysis of image recognition task under different network conditions.}
            \begin{tabular}{c|c|c|c|c}
                \hline
                \multirow{2}{*}{\textbf{Network}} & \multicolumn{4}{c}{\textbf{Normalized Overhead (\%)}} \\ \cline{2-5} 
& \textbf{System} & \textbf{Device} & \textbf{Transmission} & \textbf{Cloud} \\
                \hline
                WiFi & 0.063\% & 26.69\% & 66.89\% & 6.35\% \\
                \hline
                5G & 0.15\% & 97.30\% & 2.36\% & 0.19\% \\
                \hline
                4G & 0.21\% & 99.75\% & 0.00\% & 0.00\% \\
                \hline
            \end{tabular}
            \label{tab:overhead}
\end{table}

\section{Related Work}\label{sec:Related Work}
Most efforts in accelerating ViT model inference are concentrated on model optimization. 
Optimization methods, such as knowledge distillation\cite{hao2022learning}, pruning\cite{pan2021iared,tang2022patch,xu2022evovit,yu2022width,zheng2022savit}, quantization\cite{li2023psaqvit}, neural architecture 
search\cite{you2022shiftaddnas}, 
and designing lightweight networks\cite{yang2022lite} have been developed to optimize the inference phase of ViT models. These methods aim to reduce the computational complexity and memory demands of the model, leading to faster inference. 
However, these models and techniques only work in local device environments. 
They cannot be directly applied to the collaborative inference setting and do not fully exploit the potential of resource-rich cloud servers to accelerate inference further. 

Cloud-device collaborative inference enables the partitioned deployment of the model between the device and cloud in CNN architectures\cite{kang2017neurosurgeona,song2018insitu,jeong2018computation,hu2019dynamic,zeng2019boomerang,li2020edge,laskaridis2020spinna,ren2021finegrained,yang2022cnnpc,zhang2021cloudedgeb}.
Fundamentally, the benefit of cloud-device collaboration for inference acceleration relies on data reduction during the inference phase, leading to a shorter communication latency.  Unlike the typical CNN architecture, the unique transformer architecture lacks this property, rendering the existing cloud-device collaboration efforts inapplicable. 
In our work, we introduce the first cloud-device collaborative inference framework for emerging ViTs. 

\section{Conclusion}\label{sec:Conclusion}
In this work, we present Janus, a cloud-device collaborative 
computing system for low-latency ViT inference over dynamic networks. Janus satisfies the stringent latency requirement and delivers high accuracy through innovative adaptation of token pruning techniques and meticulously designed pruning and model splitting policies. The design perfectly captures the characteristics of the underlying computing infrastructure and the intrinsic properties of ViT models. Furthermore, we develop a lightweight profiler to accurately forecast computing latency across various candidate points. Leveraging this insight alongside the tailored pruner and splitter, we propose an efficient scheduling policy to realize collaborative inference with low computing complexity.  
Extensive evaluations based on real-world devices and network scenarios demonstrate the effectiveness of Janus in achieving low latency and maintaining high accuracy. We believe that the development of Janus not only reveals new opportunities in model-aware video analytics systems but also significantly impacts the future serving stack of emerging transformer-based AI applications.

\bibliographystyle{IEEEtran}
\bibliography{IEEEabrv,references}

\end{document}